\documentclass[manuscript]{aastex}
\usepackage{amsmath}

\begin{document}
\title{Toward Standard Data Production for Magnetic Field Observations at Huairou Solar Observing Station}
\author{Liu, S.$^{1}$, Yang, X$^{1}$., Zhao, C.$^{1}$, and Lin G.H.$^{1}$}
\altaffiltext{1}{key Laboratory of Solar Activity, National
Astronomical Observatory, Chinese Academy of Sciences, Beijing, China\\}
\email{lius@nao.cas.cn}
\begin{abstract}

The routine solar observations are available at Huairou Solar Observing Station (HSOS) since 1987. The data storage medium and format at HSOS experienced several changes, so there exist some inconveniences for solar physicist. This paper shows that the observations data of HSOS are further processed both for storage medium and format toward
international standards, in order to explore HSOS observations data for scientific research.

\end{abstract}


\section{INTRODUCTION}

Solar magnetic field measurements contribute to the understanding of solar activities. Thus, the magnetic field with the unified standard are needed.
At present, the Flexible Image Transport System (FITS) data format be regarded as a standard data production.
FITS has the experience from the definition to the several adjustments 
which make FITS toward a more scientific and reasonable status
\citep{2010A&A...524A..42P}.
World coordinate systems (WCSs) that can transform a position in a data array to a physical coordinate system are integrated in FITS, which facilitate scientists to use the observations data. There exit some regulations in WCSs standards, such as the unified key words, should be abode to make it convenient for researchers to use 
\citep{2002A&A...395.1061G,2006A&A...449..791T}.

Solar Magnetic Field Telescope (SMFT) at Huairou Solar Observing Station (HSOS), which located at north shore of Huairou reservoir (Latitude:41.3, Longitude:116.6),
is a main solar magnetic field observation instrument in the world.
SMFT operated more than 20 years to observe magnetic field of solar photosphere and chromosphere,
and it made outstanding contributions in the research of solar photospheric magnetic field.
During its operation, the observation data of the storage medium, the storage format and data read method has undergone a few changes.

The promise of data-driven and science-driven that are based on multiband or multiple solar activity cycles is now being recognized broadly, and there is growing enthusiasm for the notion of ¡°Big Data¡±. In fact, The white paper of VSO (Virtual Solar Observatory) presented needs of data fusion. But there is currently a wide gap between its potential and its realization. Most of the reasons are caused by standard, specification and the speciality of data process.

This paper will introduce simply the magnetic field observations by SMFT, and shows that magnetic field observations toward standard data production.

\section{MAGNETIC FIELD OBSERVATIONS}
The routine solar magnetic field of photosphere observations are available at HSOS since 1987,
the observations are carried out by SMFT that is a 35-cm vacuum refraction solar telescope.
 It works at FeI 5324.19\AA ~for photosphere vector magnetic field, H$\beta$ 4861.34\AA ~for chromosphere longitudinal magnetic field.
Vector magnetic field is reconstructed from Stokes parameters (I, Q, U and V). 
The conventional linear calibration is used to calibrate magnetic field under the weak-field approximation.
\begin{equation}
B_{L}=C_{L}V ,\\
\end{equation}
\begin{equation}
B_{T}=C_{T}(Q^{2}+U^{2})^{1/4}, \\
\end{equation}
\begin{equation}
\theta=arctan(\dfrac{B_{L}}{B_{\bot}}),\\
\end{equation}
\begin{equation}
\phi=\dfrac{1}{2}arctan(\dfrac{U}{Q}),\\
\end{equation}
where $B_{L}$ and $C_{L}$ are the line-of-sight magnetic field and the corresponding calibration coefficients,
$B_{T}$ and $C_{T}$ are transverse magnetic field and and the corresponding calibration coefficients.
$\theta$ and $\phi$ are the inclination and azimuth of field.
The magnetic field calibration for this observation data can referee papers of
\citep{1996ArBei..28...31W,2004ChJAA...4..365S},

\section{HISTORICAL OBSERVATIONS DATA}
There are a few changes of data storage medium, format and read method during SMFT operation.
At present, the
observations data is stored in the form of electronic one.
Because of the computer technology, the length of filename of early data is restricted.
The early data contain directly Stokes images which are deduced by two polarization component either linear or circular correspondingly.
After April 2009, two components of linear or circular polarization are respectively recorded in data, the Stokes images are deduced by the two components, and for data alignment one polarization component can be used to compensated observed time difference.
The early data processed and stored as DAT, and the header contain the limited information of wavelength, observations time,
weather, coordinates, stack frames and active region number. After 2001, the observations data is stored in hard disk as FITS directly,
and the information about observations contained in header is more plentiful, however some key words in header deviate from unified standards.

\section{TOWARD STANDARD OBSERVATIONS DATA}

\subsection{DAT TRANSFORM TO FITS}
Transforming data form DAT format to FITS format is the necessary intermediate process that targets to get finally the standard data.
For SMFT observations data before 2001 has already stored in disk as DAT format, so transform these to FITS is needed.
To do this, the original data contained in DAT is transform/copy to FITS, while the data information contained in DAT is
remade as FITS header regularly. The filename is renamed to make sure that it can give information of measurements, observation date and time, active region number.

\subsection{CALIBRATION LONGITUDINAL MAGNETIC FIELD}
Under the weak-field approximation and a series of related
works
\citep{1996ArBei..28...31W,2004ChJAA...4..365S},
all routine Stokes V images are calibrated to longitudinal magnetic field.
And renamed the filename to contain information of measurements, observation date and time, active region number.
The calibration coefficient of longitudinal field is 8381. The FITS header is standardized and the pixel size. 
At last, the calibrated longitudinal field written/stored as FITS.

\subsection{RECONSTRUCTION VECTOR MAGNETIC FIELD}

\begin{figure}

   \centerline{\includegraphics[width=1\textwidth,clip=]{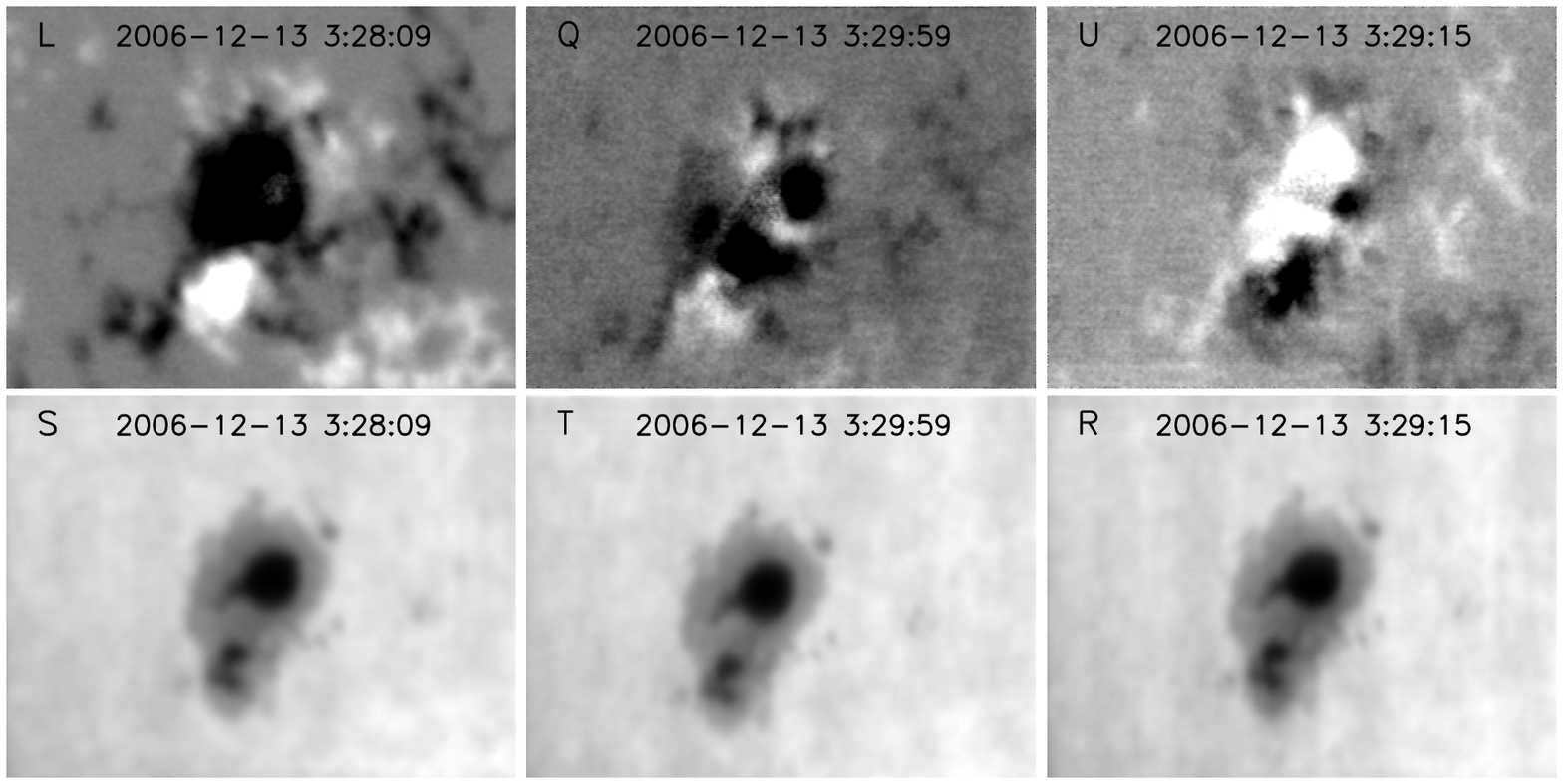}}

   \caption{The images of Stoke-Q/U/V signal and the corresponding filter images of T/R/S.} \label{Fig1}
\end{figure}

\begin{figure}

   \centerline{\includegraphics[width=1\textwidth,clip=]{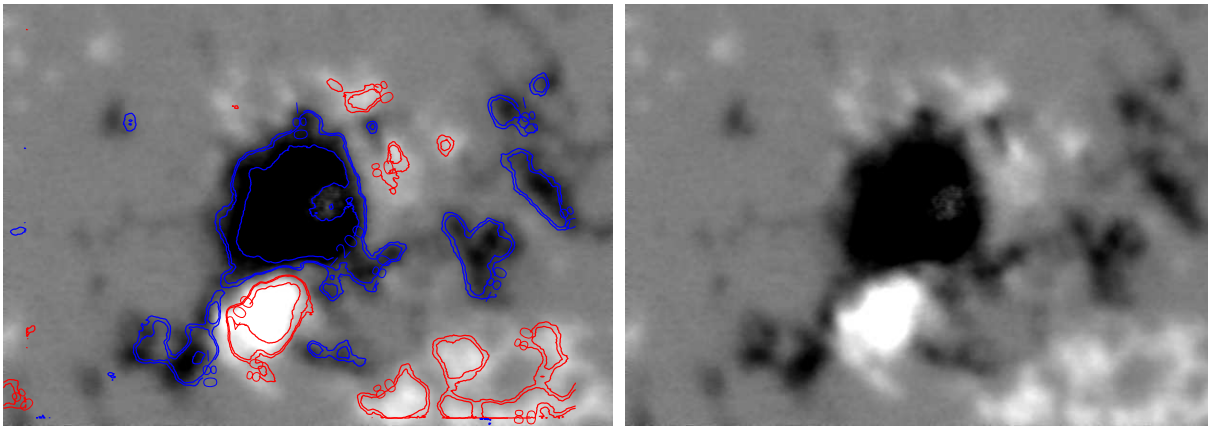}}

   \caption{The reconstructed vector magnetic field, here the transverse field are labeled by the short lines.} \label{Fig2}
\end{figure}

To observe solar active region vector magnetic field is the original intention of SMFT.
For SMFT, the fixed field and quasi simultaneous observations of a group Stokes Q/U/V are used to deduce vector magnetic field,
which is the process that contain magnetic field calibration and spatial alignment of magnetic field components.
When SMFT observe Stokes (/Q/U/V), the corresponding filter images (T/R/S) are simultaneously observed,
which are used to compensate for the time differences during the observations of Q, U and V.
The reconstruction of vector magnetic field are as follows: 
Firstly, one group of Stokes (/Q/U/V) and filter images (T/R/S) should be found with a specific interval required,
here the routine time differences between the components of Q/T, U/R and L/S are 15 mins (for some special cases this time
interval can be modified reasonably).
Secondly, the filter images of T, R and S are spatially aligned through feature correlation basing Fourier transformation
technique.
Basing on the spatial movements (only a few pixels normally) of individual filter images,
the Stokes Q, U and V are aligned under assumption that Q/U/V and T/R/S are aligned spontaneously.
Thirdly, the weak-field approximation is use to calibrate magnetic field (equations 1-2).
Additionally, the transverse field ($B_{t}$) is  decomposed to two components on the surface basing on equation (4),
the two components of $B_{t}$ are indicated by $B_{x}$ and $B_{y}$.
In this process, the the calibration coefficient of longitudinal/transverse field are 8381/6790, respectively.
Examples are shown by Fig\ref{Fig1} and Fig\ref{Fig2},
Fig\ref{Fig1} shows Stoke-Q/U/V signal and the corresponding filter images of T/R/S. Fig\ref{Fig2} shows
the reconstructed vector magnetic field, the gray-scale image is longitudinal field and the lines indicate
the transverse field.
Fourthly, write/store the calibrated vector magnetic field as FITS
and reorganize original observations information as header stored in FITS.

\subsection{STANDARDIZATION FILE INFORMATION}
The above standardization processes are all
accompanied by the processes of file information standardization. At first, the FITS file processed should be renamed with filename that can show the necessary observations information, such as measurements, observation date, observation time and active region number. 
Then, the observations information included in DAT file and header of original observed FITS is rearranged as
new header written in FITS file. Key words in header are adjusted toward standardization, such as \emph{T\_OBS, OBS\_ATE, OBS\_TIME, XSCALE} and \emph{YSCALE}. 
To standardize observations data, the NOAA number that can be matched and recognized is added in the header and
the one type of WCS is provided in FITS header.

\section{CONCLUSIONS}
The photosphere magnetic field observations data obtained from SMFT at HSOS is standardized to try to satisfy international standards. The processes included the transfer of storage media,
the conversion of data format, calibration of longitudinal magnetic field, reconstruction of vector magnetic field, standardizations of data information  and filename. At mean time, WCS deduced from original information is added in FITS header. Additionally, some standardizations that do not carry on or that come out in future will be applied to SMFT observations.

\section{Acknowledgements}
It was partly supported by the Grants: : 2014FY120300, 2011CB811401, KJCX2-EW-T07, 11203036, 11221063, 11373040, 11178005, 10673016, 10778723 and 11178016, the Key Laboratory of Solar Activity National Astronomical Observations, Chinese Academy
of Sciences.



\end{document}